\documentclass[aps,notitlepage, floatfix, prr, twocolumn]{revtex4-2}

\usepackage[colorlinks=true, allcolors=blue, bookmarks=false]{hyperref}
\usepackage{amsfonts, amsmath, amssymb}
\usepackage{graphicx}
\usepackage{bbm}

\DeclareMathOperator{\diag}{diag}

\begin{document}

\title{Efficient circular Dyson Brownian motion algorithm}

\author{Wouter Buijsman} 

\email{buijsman@pks.mpg.de}

\affiliation{Department of Physics, Ben-Gurion University of the Negev, Beer-Sheva 84105, Israel}

\affiliation{Max Planck Institute for the Physics of Complex Systems, N\"othnitzer Str. 38, 01187 Dresden, Germany}

\date{\today}

\begin{abstract}
Circular Dyson Brownian motion describes the Brownian dynamics of particles on a circle (periodic boundary conditions), interacting through a logarithmic, long-range two-body potential. Within the log-gas picture of random matrix theory, it describes the level dynamics of unitary (``circular") matrices. A common scenario is that one wants to know about an initial configuration evolved over a certain interval of time, without being interested in the intermediate dynamics. Numerical evaluation of this is computationally expensive as the time-evolution algorithm is accurate only on short time intervals because of an underlying perturbative approximation. This work proposes an efficient and easy-to-implement improved circular Dyson Brownian motion algorithm for the unitary class (Dyson index $\beta = 2$, physically corresponding to broken time-reversal symmetry). The algorithm allows one to study time evolution over arbitrarily large intervals of time at a fixed computational cost, with no approximations being involved.
\end{abstract}

\maketitle

\section{Introduction}
Brownian motion describes the stochastic dynamics of microscopic particles in a thermal environment \cite{Uhlenbeck30, Wang45}. It connects a broad variety of topics, including thermal physics, hydrodynamics, reaction kinetics, fluctuation phenomena, statistical thermodynamics, osmosis, and colloid science \cite{Philipse18}. Brownian motion is intimately related to random matrix theory, which plays a key role in the understanding of quantum statistical mechanics and quantum chaos \cite{Mehta91, Haake10, DAlessio16}. Random matrices have eigenvalue statistics that typically can be studied using the so-called log-gas picture \cite{Dyson62, Forrester10}. For matrices with real eigenvalues, the joint probability distribution $P$ of the eigenvalues is then written as a Boltzmann factor
\begin{equation}
P = \frac{1}{\mathcal{Z}} e^{- \beta H},
\label{eq: log}
\end{equation}
where $\mathcal{Z}$ is a normalization constant that has the interpretation of a partition function, and $\beta > 0$ is a parameter known as the Dyson index that has the interpretation of an inverse temperature. The Hamiltonian $H$ describes a collection of classical massless particles on a line (the eigenvalues) repelling each other over long ranges through a logarithmic two-body potential, held together by a confining background potential. 

The log-gas picture describes long-range interacting particles. It has been found, for example, to accurately describe the level statistics across the many-body localization transition \cite{Buijsman19}. As the Hamiltonian in Eq. \eqref{eq: log} does not contain a kinetic term, the particles obey non-trivial dynamics. The equilibrium as well as the non-equilibrium dynamics of the particles (``level dynamics'') are described by a phenomenon referred to as Dyson Brownian motion \cite{Dyson62-2, Dyson72}. Dyson Brownian motion turns out to provide a good description rather generically when long-range interactions are involved. As such, these dynamics (as well as the corresponding stochastic evolution of the eigenstates \cite{Bourgade17, Benigni21}) have found applications in studies on, for example, disordered systems \cite{Beenakker93, Narayan93, Shukla99, Shukla05}, random matrix models \cite{Facoetti16, VonSoosten18, Kutlin21, Buijsman22}, many-body localization \cite{Serbyn16, Monthus16}, quantum information dynamics \cite{Fidkowski21, Schomerus22, Bulchandani24}, and cosmological inflation \cite{Battefeld15, Wang16, Dias16, Pedro17}. 

Circular Dyson Brownian motion for unitary (``circular") matrices describes Dyson Brownian motion on a circle (periodic boundary conditions) and without background potential. A common scenario is that one wants to know about an initial configuration evolved over a certain interval of time, without being interested in the intermediate dynamics. Dyson Brownian motion can be evolved over a time interval of arbitrary length at a fixed computational cost, with no approximations being involved (see below for a more detailed explanation). Circular Dyson Brownian motion, however, requires extensively many evaluations over small intermediate intervals because of a perturbative approximation underlying the time-evolution algorithm. Circular Dyson Brownian motion is thus a process that is computationally expensive to simulate, which moreover is subject to a loss of accuracy with progressing time. Despite significant recent \cite{Pandey17, Forrester24} and less recent \cite{Pandey91, Forrester96, Forrester98, Nagao03} analytical progress on circular Dyson Brownian motion out of equilibrium, improved numerical capabilities are thus desired.

This work proposes an improved, easy-to-implement circular Dyson Brownian algorithm for the unitary class (Dyson index $\beta = 2$, corresponding to systems with broken time-reversal symmetry). The algorithm does not require intermediate evaluations, and thus operates at dramatically lower computational cost compared to the currently used algorithm. Moreover, it does not involve approximations, and is thus not subject to a loss of accuracy with progressing time. In short, it constructs the desired unitary matrices by orthonormalizing the columns of certain non-Hermitian matrices for which the elements perform Brownian motion. Similar to Dyson Brownian motion for Hermitian matrices, this Brownian motion process can be time evolved at a computational cost independent of the length of the time interval.

\section{Dyson Brownian motion for Hermitian and unitary matrices}
One distinguishes between orthogonal ($\beta = 1$), unitary ($\beta = 2$), and symplectic ($\beta = 4$) random matrix ensembles \cite{Dyson62}. These names reflect the type of transformations under which the ensembles remain invariant. Physically, the type of invariance determines the behavior of a system under time reversal. For example, the orthogonal class correspond to time-reversal systems, whereas the unitary class correspond to systems with broken time-reversal symmetry. This section considers the unitary class, which is arguably the most convenient one.

Let $H(t)$ be an $N \times N$ Hermitian matrix with elements depending on time $t$ \cite{Dyson62-2}. The initial condition $H(0)$ can be either random or deterministic. Dyson Brownian motion for Hermitian matrices of the unitary class is a stochastic process described by
\begin{equation}
H(t + dt) = H(t) + \sqrt{dt} M,
\label{eq: H-evolution}
\end{equation}
where the time step $dt$, in order for the eigenvalue dynamics to obey Dyson Brownian motion, is supposed to be small enough such that the eigenvalues of $H(t + dt)$ can be obtained accurately by second-order perturbation theory. Here, $M$ is a sample from the Gaussian unitary ensemble that is resampled at each evaluation. An $N \times N$ matrix $M$ sampled from the Gaussian unitary ensemble can be constructed as
\begin{equation}
M = \frac{1}{2} ( A + A^\dagger ),
\end{equation}
where $A$ is an $N \times N$ matrix with complex-valued elements $A_{nm} = u_{nm} + i v_{nm}$ with $u_{nm}$ and $v_{nm}$ sampled independently from the normal distribution with mean zero and variance $1/2$.

Let $\tilde{M} = U^\dagger (t) M U(t)$, where the time-dependent unitary matrix $U(t)$ is chosen such that it diagonalizes $H(t)$. The Gaussian unitary ensemble is invariant under unitary transformations, meaning that $\tilde{M}$ can be replaced by a new sample from the Gaussian unitary ensemble. The increments $d \lambda_n(t) = \lambda_n(t + dt) - \lambda_n(t)$ of the eigenvalues $\lambda_n (t)$ when evolving from time $t$ to $t + dt$ obey
\begin{equation}
d \lambda_n(t) = \sqrt{dt} \tilde{M}_{nm} + \sum_{m \neq n} \frac{dt |\tilde{M}_{nm}|^2}{\lambda_m(t) - \lambda_n(t)},
\label{eq: lambda-evolution}
\end{equation}
where terms of order three and higher have been ignored. It can be shown that this time-evolution indeed describes a Brownian motion process, for example by writing down the corresponding Fokker-Planck equation. For $t \to \infty$, $H(t)$ converges to a (scaled) sample from the Gaussian unitary ensemble irrespective of the initial condition $H(0)$.

Dyson Brownian motion can also be studied for unitary matrices \cite{Dyson62-2, Pandey91}. Let $Q(t)$ be an $N \times N$ unitary matrix with time-dependent elements. Similar to the above, the initial condition $Q(0)$ can be either random or deterministic. Circular Dyson Brownian motion for the unitary class is generated by
\begin{equation}
Q(t + dt) = Q(t) e^{i \sqrt{dt} M},
\label{eq: Q-evolution}
\end{equation}
where again $M$ is an $N \times N$ sample from the Gaussian unitary ensemble that is re-sampled at each evaluation. For small enough $dt$, the matrix exponent can be approximated by the first-order expansion $\mathbbm{1} + i \sqrt{dt} M$, which is invariant under unitary transformations (the second and higher-order terms are not). The matrix $Q(t+dt)$ is thus obtained by applying infinitesimal orthonormality-preserving random rotations on the columns of $Q(t)$. ``Random'' here means that rotations in each direction are equally likely, which agrees with the observation that $\mathbbm{1} + i \sqrt{dt} M$ is invariant under unitary transformations.

Let $\tilde{M} = U^\dagger (t) \, M \, U(t)$ with the time-dependent unitary matrix $U(t)$ chosen such that it diagonalizes $Q(t)$. As before, $\tilde{M}$ can be replaced by a new sample from the Gaussian unitary ensemble. Circular Dyson Brownian motion of the eigenvalues $e^{i \theta_1 (t)}, e^{i \theta_2 (t)}, \dots, e^{i \theta_N (t)}$ entails that the increments $d \theta_n(t) = \theta_n(t + dt) - \theta_n(t)$ of the eigenphases $\theta_n(t)$ when evolving from time $t$ to $t +dt$ are given by
\begin{equation}
d \theta_n(t) = \sqrt{dt} \tilde{M}_{nm} + \sum_{m \neq n} \frac{dt |\tilde{M}_{nm}|^2}{2 \tan \tfrac{1}{2} [\theta_m(t) - \theta_n(t)]},
\label{eq: theta-evolution}
\end{equation}
where terms of order three and higher have been ignored. Equations \eqref{eq: lambda-evolution} and \eqref{eq: theta-evolution} describe similar dynamics on a microscopic scale since $2 \tan(x / 2) = x + \mathcal{O}(x^2)$. For $t \to \infty$, $Q(t)$ converges to a sample from the circular unitary ensemble irrespective of the initial condition $Q(0)$.

\section{The algorithm}
The Gaussian random matrix ensembles have the property that the sum of $n$ independent samples is a sample again, although with a prefactor $\sqrt{n}$. Equation \eqref{eq: H-evolution} and its equivalents for the orthogonal and symplectic classes thus do not require the time step $dt$ to be small. This implies that numerically obtaining $H(T)$ from $H(0)$ can be done in a single instance, at a computational cost independent of $T$. Equation \eqref{eq: Q-evolution} for the evolution of unitary matrices does not allow for a similar argument since $e^{A} e^{B} \neq e^{A + B}$ when $A$ and $B$ do not commute. Time-evolution for unitary matrices can thus naively only be accomplished by subsequently evolving over infinitesimal time intervals. Equation \eqref{eq: Q-evolution} moreover is subject to a loss of accuracy with progressing time as it describes the desired dynamics only up to first order.

The starting point in establishing an improved algorithm is the observation that a random unitary matrix (circular unitary ensemble) can be obtained by orthonormalizing a set of random vectors \cite{Eaton07, Mezzadri07}. Let $A$ be an $N \times N$ matrix with elements $A_{nm} = u_{nm} + i v_{nm}$ with $u_{nm}$ and $v_{nm}$ sampled independently from the normal distribution with mean zero and unit variance. Such a matrix is known as a sample from the Ginibre unitary ensemble \cite{Forrester10, Burda14}. The QR decomposition
\begin{equation}
A = Q R
\end{equation} 
decomposes $A$ in a unitary matrix $Q$ and an upper-triangular matrix $R$ with real-valued diagonal elements. This decomposition is not unique. It can be made unique by fixing the signs of the diagonal elements of the upper-triangular matrix, for example, by requiring them to be non-negative. Let 
\begin{equation}
\Lambda = \diag \left( \frac{R_{11}}{| R_{11} |}, \frac{R_{22}}{| R_{22} |}, \dots, \frac{R_{NN}}{| R_{NN} |} \right).
\end{equation}
Then, $Q \to Q \Lambda$ and $R \to \Lambda R$ is the QR decomposition with the upper-triangular matrix having non-negative diagonal entries. One can prove that the resulting unitary matrices $Q$ obey the distribution of the circular unitary ensemble. Algorithmically, such unitary matrices are obtained by performing Gram-Schmidt orthonormalization (discussed below) on the columns of $A$. A sample from the circular unitary ensemble can thus be obtained by orthonormalizing a set of random vectors.

Let $U(dt)$ be an $N \times N$ unitary matrix with time-dependent elements. The goal is to express $Q(t + dt)$ of Eq. \eqref{eq: Q-evolution} as
\begin{equation}
Q(t+dt) = Q(t) \, U(dt),
\label{eq: U}
\end{equation}
where $dt$ is not necessarily small. Equation \eqref{eq: Q-evolution} indicates that the dynamics of $Q(t)$ are generated by orthonormality-preserving random rotations of the columns. Thus, $U(dt)$ interpolates between an identity matrix ($dt = 0$) and a sample from the circular unitary ensemble ($dt \to \infty$) in a way such that $U(dt)$ is invariant under unitary transformations. In other words, it generates a finite orthonormality-preserving random rotation of the columns of $Q(t)$. Generalizing the above algorithm generating random unitary matrices, consider the QR decomposition
\begin{equation}
\mathbbm{1} + \sqrt{d\tau} A = U(d\tau) \, R \qquad (R_{nn} \ge 0).
\label{eq: U-QR}
\end{equation}
Here, $A$ is again a sample from the Ginibre unitary ensemble. This ensemble is invariant under unitary transformations. The parameter $d\tau$ is some yet undetermined function of $dt$, which for small enough $d\tau$ will be found to be equal to $dt$. The aim is to show that $U(d\tau)$ corresponds to $U(dt)$ of Eq. \eqref{eq: U}. In Eq. \eqref{eq: U-QR}, the columns $u_n$ of $U(d\tau)$ result from Gram-Schmidt orthonormalization of the columns $m_n$ of the left-hand side,
\begin{equation}
u_n = \frac{v_n}{|| v_n ||}, \qquad v_n = m_n - \sum_{m=1}^{n-1} (u_m \cdot m_n) u_m.
\end{equation}
In words, the $n$-th column is obtained by substracting the projections on the first $n-1$ columns, followed by normalization. Columns with a higher index undergo more substractions than columns with lower indices. For $N d\tau \ll 1$, these substractions do not significantly alter the directions of the columns, which are then rotated randomly since $\mathbbm{1} + i \sqrt{d\tau} A$ is invariant under unitary transformations. As the columns of $U(d\tau)$ are rotated randomly, $U(d\tau)$ corresponds to $U(dt)$ of Eq. \eqref{eq: U} for a proper choice of $d\tau$, provided that $N d\tau \ll 1$.

The limitation on the maximum value of $d\tau$ can easily be overcome by adapting a different, appropriate, orthonormalization procedure. L\"owdin symmetric orthonormalization is a procedure for which the columns are treated symmetrically, that is, the outcome is independent of the ordering \cite{Lowdin50}. For $M$ denoting some matrix, consider the singular value decomposition 
\begin{equation}
M = U_1 \Sigma U_2^\dagger.
\end{equation}
Here, $U_{1,2}$ are unitary matrices and $\Sigma$ is a diagonal matrix with real-valued nonnegative entries. L\"owdin symmetric orthonormalization gives the unitary matrix $U = U_1 U_2^\dagger$, which can be shown to be optimal in the sense that the distance
\begin{equation}
d = \sum_n \bigg| \bigg| \frac{m_n}{||m_n||} - u_n \bigg| \bigg|
\end{equation}
between the columns $m_n$ of $M$ and $u_n$ of $U$ acquires the minimal possible value \cite{Pratt56}. This invites to consider the SVD-decomposition
\begin{equation}
\mathbbm{1} + \sqrt{d\tau} A = U_1 \Sigma U_2^\dagger, \qquad U(d\tau) = U_1 U_2^\dagger,
\label{eq: U-SVD}
\end{equation}
with $A$ denoting a sample from the Ginibre unitary ensemble. If $M \to U$ by L\"owdin symmetric orthonormalization, then $V^\dagger M V \to V^\dagger U V$ for unitary matrices $V$. The Ginibre unitary ensemble is invariant under unitary transformations. These two facts combined guarantee the rotations generated by $U(d\tau)$ to be random. Thus, $U(d\tau)$ corresponds to $U(dt)$ of Eq. \eqref{eq: U} for a proper choice of $d\tau$, without $d\tau$ required to be small.

\begin{figure}
\includegraphics[width = \columnwidth]{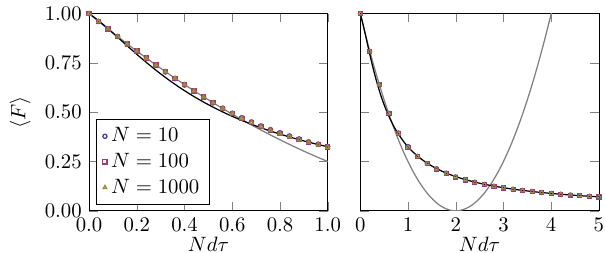}
\caption{Plots of the ensemble-averaged value of $F$ [Eq. \eqref{eq: F}] for dimensions $N = 10$, $N = 100$, and $N = 1000$ as a function of $Nd\tau$ on a small (left) and larger (right) range. The solid lines show plots of $(1 - Nd\tau / 2)^2$ and $\big( (1 + Nd\tau)^{-2} + 2 \sqrt{2} N d\tau \big)^{-1}$ in gray and black, respectively.}
\label{fig: dtau}
\end{figure}

The relation between $dt$ and $d\tau$ can be established by requiring $U(dt)$ [Eq. \eqref{eq: U}] and $U(d\tau)$ [Eq. \eqref{eq: U-SVD}] to be identically distributed. For $U(d\tau)$ [Eq. \eqref{eq: U-SVD}], let $u_1(d\tau)$ denote the first column (the choice for the first column is arbitrary), and consider the overlap
\begin{equation}
F(d\tau) = \frac{N}{N-1} \bigg( |u_1(d\tau) \cdot u_1(0)|^2 - \frac{1}{N} \bigg).
\label{eq: F}
\end{equation}
The overlap is shifted and scaled such that $F(0) = 1$ and $F(\infty) = 0$. Figure \ref{fig: dtau} shows that the ensemble average of $F$ is almost perfectly described at all times already for $N = 10$ by $F = (1 - N d\tau/2)^2$ before and $F = \big( (1 + Nd\tau)^{-2} + 2 \sqrt{2} N d\tau \big)^{-1}$ after the intersection at $N d\tau \approx 0.66211710937$. These expressions have been found empirically. Next consider $U(dt)$ [Eq. \eqref{eq: U}]. Equation \eqref{eq: Q-evolution} dictates, as can be verified numerically, that the product of two independent samples $U(dt_1)$ and $U(dt_2)$ is from the same distribution as $U(dt_1 \, dt_2)$. For $F$ defined similar as above, this means that $F(dt) = e^{-Ndt}$ since $e^{-N dt_1} e^{-N dt_2} = e^{-N (dt_1 + dt_2)}$. Equating $F(dt)$ to the piecewise expression $F(d\tau)$ introduced above gives
\begin{equation}
N dt = 
\begin{cases}
-2 \ln(1 - N d\tau / 2) & \text{if } N d\tau \le 0.662, \\
\ln \big( (1 + N d\tau)^{-2} + 2 \sqrt{2} N d\tau \big) & \text{if } N d\tau > 0.662, \\
\end{cases}
\label{eq: dt}
\end{equation}
which can be inverted numerically to find $d\tau$ as a function of $dt$. Up to first order, the approximation $dt = d\tau$ can be made.

\section{Numerical verification}
\begin{figure}
\includegraphics[width = \columnwidth]{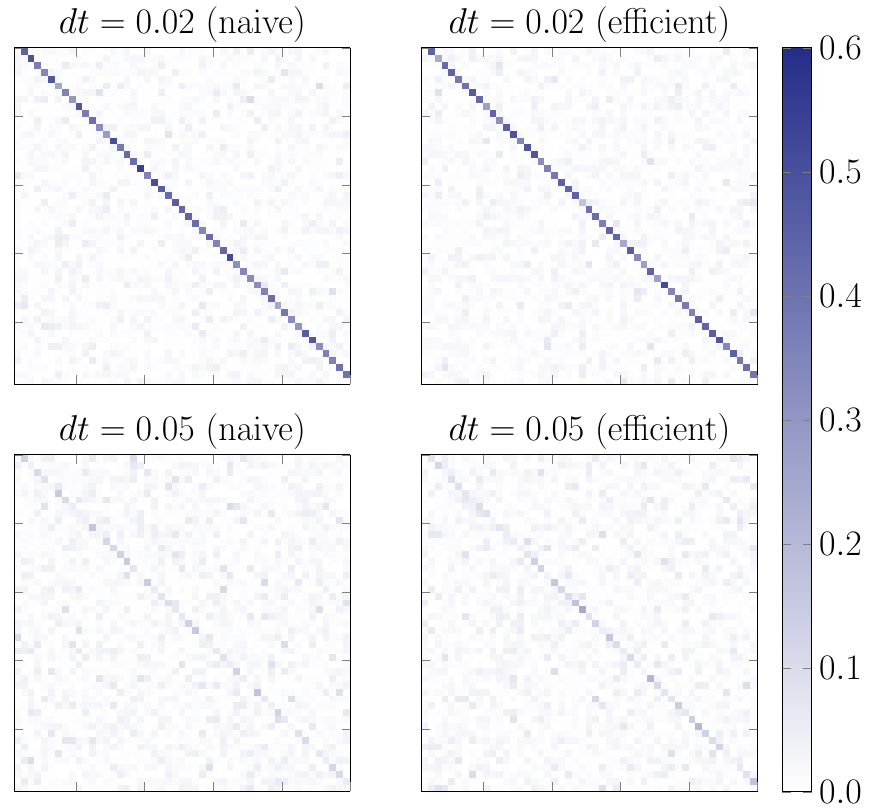}
\caption{Density plots of $|Q(dt)|^2$ obtained through Eq. \eqref{eq: Q-evolution} [``naive''] and Eqs. \eqref{eq: U}, \eqref{eq: U-SVD}, and \eqref{eq: dt} [``efficient''] for $dt = 0.02$ (top, $d\tau \approx 0.017$) and $dt = 0.05$ (bottom, $d\tau \approx 0.086$). Here, $N = 50$ and $Q(0) = \diag(1,1,\dots, 1)$.}
\label{fig: matrices}
\end{figure}

This Section provides a numerical verification of the algorithm proposed above. First, the focus is on the structure of the resulting matrices. Figure \ref{fig: matrices} shows density plots of $|Q(dt)|^2$ for matrices of dimension $N = 50$ at short ($dt = 0.02$) and longer ($dt = 0.05$) times obtained through Eq. \eqref{eq: Q-evolution} [left, ``naive''] and Eqs. \eqref{eq: U}, \eqref{eq: U-SVD}, and \eqref{eq: dt} [right, ``efficient'']. The initial condition $Q(0) = \diag(1, 1, \dots, 1)$ is taken such that $Q(dt) = U(dt)$. The values of $d\tau$ corresponding to these values of $dt$ are given in the caption. One observes that the matrices on the left and right show identical characteristics.

\begin{figure}
\includegraphics[width = \columnwidth]{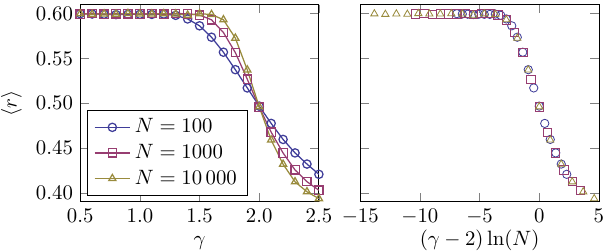}
\caption{The average ratio of consecutive level spacings $\langle r \rangle$ as a function of $\gamma$ [left] and $(\gamma - 2) \ln(N)$ [right] for the unitary equivalent of the Rosenzweig-Porter model. The data is obtained using Eqs. \eqref{eq: U}, \eqref{eq: U-SVD}, and \eqref{eq: dt}. See the main text for details.}
\label{fig: rav}
\end{figure}

A sample from the unitary equivalent of the Rosenzweig-Porter model, considered next, can be obtained as $Q(dt = N^{-\gamma})$ by taking $Q(0) = \diag(e^{i \theta_1^{(0)}}, e^{i \theta_2^{(0)}}, \dots e^{i \theta_N^{(0)}})$ with the phases $\theta_n^{(0)}$ sampled independently from the uniform distribution ranging over $[0, 2\pi)$ \cite{Buijsman22}. See Refs. \cite{Rosenzweig60, Kravtsov15} for an introduction to the Rosenzweig-Porter model and its relation to Dyson Brownian motion. Level statistics are here quantified by the average ratio $\langle r \rangle$ of consecutive level spacings \cite{Oganesyan07, Atas13}. For unitary matrices with ordered eigenphases $\theta_n$, the $n$-th ratio is defined as
\begin{equation}
r_n = \min \bigg( \frac{\theta_{n+2} - \theta_{n+1}}{\theta_{n+1} - \theta_n}, \frac{\theta_{n+1} - \theta_n}{\theta_{n+2} - \theta_{n+1}} \bigg).
\end{equation}
The average is taken over all $n$ and a large number of realizations. Wigner-Dyson level statistics are characterized by $\langle r \rangle \approx 0.600$, while Poissonian level statistics obey $\langle r \rangle \approx 0.386$. The Rosenzweig-Porter model shows a transition (at finite dimension, a crossing) from Wigner-Dyson to Poissonian level statistics at $\gamma = 2$. When plotted as a function of $(\gamma - 2) \ln(N)$, the average ratio is numerically found to be independent of $N$ (finite-size collapse) \cite{Pino19, Buijsman22}. Figure \ref{fig: rav} shows that the algorithm proposed in this work leads to the same results, and illustrates the capability of the algorithm proposed in this work to operate at large matrix dimensions (here, up to $N = 10 \, 000$). Reference \cite{Buijsman24} (Fig. 1) shows a visually indistinguishable plot obtained using Eqs. \eqref{eq: U}, \eqref{eq: U-QR} with the first-order approximation $dt = d\tau$.

\section{Conclusions and outlook}
Circular Dyson Brownian motion describes the Brownian dynamics of particles interacting through a long-range two-body potential in a one-dimensional environment with periodic boundary conditions. This work proposed an easy-to-implement algorithm [Eqs. \eqref{eq: U}, \eqref{eq: U-SVD}, and \eqref{eq: dt}] to simulate circular Dyson Brownian motion for the unitary class (Dyson index $\beta = 2$, physically corresponding to broken time-reversal symmetry). For short times $N dt \ll 1$, Eq. \eqref{eq: U-SVD} can be replaced by the computationally cheaper Eq. \eqref{eq: U-QR}, and the first-order approximation $dt = d\tau$ can be used instead of the more complicated relation \eqref{eq: dt}. The latter approach is a generalization of a commonly used algorithm generating samples from the circular unitary ensemble, proposed in Refs. \cite{Eaton07, Mezzadri07}. In contrast to the currently used circular Dyson Brownian motion algorithm [Eq. \eqref{eq: Q-evolution}], here the time step $dt$ does not have to be small, and no approximations have been involved. This allows one to study time-evolution over arbitrarily large time intervals at a computational cost independent of the length of the time interval, without loss of accuracy. In typical settings, this algorithm dramatically reduces the computational costs, thereby for example opening the possibility to perform detailed studies without the need for high-performance computing facilities.

An arguably interesting follow-up question would be how to modify the algorithm for the orthogonal and symplectic classes. From a sample $Q$ of the circular unitary ensemble, a sample $S$ from the circular orthogonal ensemble can be obtained as $S = Q^T Q$ \cite{Zyczkowski93}. It is thus tempting to hypothesize that circular Dyson Brownian motion for the orthogonal class can be simulated by the algorithm proposed in this work, and by taking the product of the transpose of the resulting unitary matrix and the resulting unitary matrix itself as the output.

Circular Dyson Brownian motion can be used to numerically generate non-ergodic unitary matrices (“unitaries”) with fractal eigenstates and a tunable degree of complexity \cite{Buijsman22, Buijsman24}. Next to what is mentioned above, this work can thus be expected to be relevant for future studies on the emergence and breakdown of statistical mechanics in the context of unitary (periodically driven) systems. It also relates to recent developments on algorithms generating random rotations \cite{Bullerjahn23}. Dyson Brownian motion recently attracted a spurge of interest in the context of the Brownian SYK model \cite{Sunderhauf19, Jian21, Agarwal22, Zhang23, Balasubramanian23, Milekhin23, Tiutiakina24}. Unitary Brownian quantum systems are of current interest in the context of Brownian quantum circuits \cite{Zhou19, Bentsen21, Vovk22, Sahu22, Jian23, Sahu24}. This work finally can be expected to provide new opportunities in the context of the non-trivial dynamics of Brownian quantum systems.

\begin{acknowledgments}
The author acknowledges support from the Kreitman School of Advanced Graduate Studies at Ben-Gurion University.
\end{acknowledgments}

\bibliography{references}

\end{document}